\documentclass[prd,twocolumn,showpcs,amsmath,amssymb,nofootinbib,preprintnumbers,balancelastpage]{revtex4}
\usepackage{epsfig}
\usepackage{amsmath}
\usepackage{bm}
\usepackage{times}
\usepackage{graphicx}
\usepackage{color}
\usepackage{slashed}
\usepackage{graphicx}
\usepackage{amsmath}

\def\bea{\begin{eqnarray}}
\def\eea{\end{eqnarray}}
\def\ba{\begin{eqnarray}}
\def\ea{\end{eqnarray}}
\def\be{\begin{equation}}
\def\ee{\end{equation}}

\begin{document}
\preprint{CALT 68-2943}

\title{ Phenomenology of heavy vector-like leptons}

\author{Koji Ishiwata and Mark B. Wise\\
\textit{California Institute of Technology, Pasadena, CA 91125, USA}\\
}
{\today}

\begin{abstract}
  We study the impact that a heavy generation of vector-like leptons
  can have on the value of the electric dipole moment of the electron,
  and the rates for the flavor violating processes $\mu \rightarrow e
  \gamma$ and $\mu \rightarrow 3e$.  The smallness of the charged
  lepton masses suggests that at least some of the Yukawa coupling
  constants of the vector-like leptons to the ordinary leptons or
  amongst themselves are small, but even with such small couplings
  experiments trying to detect these quantities are sensitive to extra
  generation lepton masses up to about $100~{\rm TeV}$.
\end{abstract}

\maketitle
\bigskip

\section{Introduction}
We have experimental evidence for three generations of quarks and
leptons.  There are powerful constraints on the possibility of an
additional standard model like chiral generation of quarks and leptons
from LHC data.  These come mostly from the constraints on the
additional quarks.  At present we don't have a good understanding of
why there are three generations and so it is worth exploring the
possible experimental consequences of additional fermions. One
possibility is an additional vector-like generation, where a standard
model like generation is paired with one of opposite chirality. For
such fermions mass terms are allowed and so the masses of such
fermions could be much larger than the electroweak (EW) scale and
furthermore the masses of the quarks could be very different from
those of the leptons. For any value of the vector-like fermion masses
adding these particles is technically natural since there are no
quadratically divergent contributions to their masses. In this paper
we explore some of the experimental consequences of vector-like
leptons (assuming either the vector-like quarks are much heavier than
the leptons or don't exist).

Vector-like leptons have already been studied extensively in the
literature. They have been added to cancel anomalies in models that
extend the gauge group to include $U(1)$ lepton number and possibly
also baryon number (see, for example, \cite{FileviezPerez:2011pt}). We
considered the impact of vector-like leptons on the properties of the
Higgs boson under the assumption that almost all of their masses come
from their couplings to the Higgs boson~\cite{Ishiwata:2011hr}.  In
Ref.~\cite{Kannike:2011ng}, the contribution of additional leptons to
muon $g-2$ was studied. They considered several models, which consist
of extra leptons with different EW charges, and derived the Wilson
coefficient for $\bar{\mu} \sigma^{\mu\nu}\mu F_{\mu\nu}$. Here
$F_{\mu\nu}$ is field strength of the electromagnetic field. Bounds
from precision data were also discussed because the extra leptons
induce dimension six operators, such as $(H^\dagger D_\mu \tau^a
H)(\bar{L}\gamma^\mu \tau^a L)$, $(H^\dagger D_\mu H)
(\bar{L}\gamma^\mu L)$ and $(H^\dagger D_\mu H) (\bar{e}_R\gamma^\mu
e_R)$.  Here $H=(H^+,H^0)^T$ is the Higgs doublet, $L=(\nu_L,e_L)^T$ and
$e_R$ are the lepton doublets and singlets (flavor indices are
omitted), $\tau^a$ is a Pauli matrix and $D_\mu$ denotes the covariant
derivative. Constraints from precision EW measurements were also
discussed in
Refs.~\cite{delAguila:2008pw,Joglekar:2012vc,Kearney:2012zi}. In
Refs.~\cite{Joglekar:2012vc,Kearney:2012zi}, the impact of additional
vector-like matter on $h\rightarrow \tau\tau,~bb,~\gamma\gamma$ was
studied. The impact on $h\rightarrow \mu\mu$ and the renormalization
group equations for the Higgs quartic coupling, gauge couplings, and
production rate of the Higgs boson at the LHC, were also studied.
Ref.~\cite{Dermisek:2013gta} is similar.  In
Ref.~\cite{Harnik:2012pb}, flavor violating processes ($\tau
\rightarrow \mu \gamma$, $\tau \rightarrow e\gamma$, $\mu \rightarrow
e \gamma$, $\tau \rightarrow 3\mu$, $\tau \rightarrow 3e$, $\mu
\rightarrow 3e$), magnetic and electric dipole moments,
$\mu\rightarrow e$ conversion, and flavor violating Higgs decay were
studied from the effective theory point of view. They carefully
pointed out regions of parameter space which avoid a tuning of lepton
Yukawa couplings. See also Ref.~\cite{McKeen:2012av} which studied
Higgs to dilepton decay and CP and lepton flavor violating
process.

In this paper we focus on weakly coupled vector-like leptons in the
region of parameter space where they are heavy, {\it i.e.}, masses
significantly above one TeV. Although they change Higgs properties by
a very small amount and are not constrained by precision EW
measurement, we find that vector-like leptons with masses up to about
$100~{\rm TeV}$ can contribute at the present experimental limit to
lepton flavor violating muon decay and the electric dipole moment of
the electron without unnatural assignments for their Yukawa coupling
constants.  Even though our calculation of the amplitude for $\mu
\rightarrow e \gamma$ exists in the literature (see
Refs.~\cite{Kannike:2011ng,Dermisek:2013gta,delAguila:1982yu,delAguila:1982fs})
and its enhancement due to mass mixing of the vector-like fermions has
been noted previously, we do address a few new issues including the
tree-level decay rate for $\mu \rightarrow 3e$, the electric dipole
moment of the electron and the one loop correction to mass matrix for
the light leptons. Furthermore our focus on the multi-TeV regime is
different from most of the previous literature and this allows for
simplification of the formulas for dipole moments and the amplitudes
for flavor violating decays of charged leptons.

One goal of this paper is to explore whether the enhancement in the
amplitudes for the chirality flipping radiative processes $\mu
\rightarrow e \gamma$ and the electric dipole moment of the electron,
arising from heavy vector-like lepton mixing, results in sensitivity
to much higher mass scales for the new vector-like leptons. We find
that this is not the case if naturalness in the mass matrix for the
standard model charged leptons is imposed.\footnote{A similar
  situation occurs with lepton flavor violation induced at one loop by
  scalar leptoquarks, see Ref.~\cite{Arnold:2013cva} } We also stress
that for equal branching ratios $\mu \rightarrow 3e$ is sensitive to
higher masses for the vector-like leptons than $\mu \rightarrow e
\gamma$.

\section{The Model}
\label{sec:model}

We add to the standard model a fourth generation of leptons, that have
the same quantum numbers under the gauge group as the standard model
leptons.  We call them $L_L^{\prime}$, $e_R^{\prime}$ and
$L_R^{\prime\prime}$, $e_L^{\prime\prime}$ where
$L_L^{\prime}=(\nu_L',e_L')^T$ and
$L_R^{\prime\prime}=(\nu_R'',e_R'')^T$. Here $L_L^{\prime}$,
$e_R^{\prime}$ is another copy of a standard model lepton generation,
while $L_R^{\prime\prime}$, $e_L^{\prime\prime}$ is the one with the
opposite chirality. Hence there is no gauge anomaly caused by these
extra leptons. 
The new terms in the Lagrangian density that involve these additional
leptons are
\begin{eqnarray}
{\cal L}&=&-M_L\bar{L}'_L L''_R -M_E\bar{e}'_R e''_L
-h'_E\bar{L}'_L H e'_R \\
&& -h''_E\bar{L}''_R H e''_L- {\lambda}^i_E\bar{L}'_L H e^i_R 
-{\lambda}^i_L\bar{L}^i_L H e'_R 
+h.c. .\nonumber
\label{eq:L}
\end{eqnarray}
$L_L^i,e_R^i$ ($i=e,\mu,\tau$) are mostly the standard model leptons.
In addition there are the gauge invariant kinetic terms involving
these leptons. We have chosen a basis where terms such as
$\bar{L}_R''L_L^i$ and $\bar{e}_L''e_R^i$ are rotated away. In that
basis we write the Yukawa interactions of the ordinary generations as
\begin{equation}
\label{smyuk}
 {\cal L}_{Y}=-Y_e^{ij} {\bar L_L}^{i}H e_R^{j} + h.c..
\end{equation}
Introducing the four component fields, $e_1=(e'_L,e''_R)^T$,
$e_2=(e''_L,e'_R)^T$ and $\nu_{1}=(\nu'_L,\nu''_R)^T$, the
Lagrangian density becomes,
\begin{eqnarray} 
&&{\cal L}=
  -M_L \bar{e}_{1L}e_{1R}  -M_L \bar{\nu}_{1L}\nu_{1R} -M_E\bar{e}_{2R}e_{2L} 
\nonumber \\ && 
- h'_E (H^0\bar{e}_{1L} + H^+\bar{\nu}_{1L} ) e_{2R}
- h''_E(H^0\bar{e}_{1R} + H^+\bar{\nu}_{1R} ) e_{2L}
\nonumber \\ &&
- {\lambda}^i_E (H^0\bar{e}_{1L}  + H^+\bar{\nu}_{1L} ) e^i_{R}
  - {\lambda}^i_LH^0\bar{e}^i_{L} e_{2R}
+h.c.. 
\end{eqnarray}

We are primarily interested in the processes $\mu \rightarrow e
\gamma$, $\mu \rightarrow 3e$ and the electric dipole moment (EDM) of
the electron.  One reason is that the experimental bounds for the
processes are very stringent, {\it i.e.}, the branching fraction for
$\mu \rightarrow e \gamma$ and $\mu \rightarrow 3e$ should
satisfy~\cite{Adam:2011ch,Bellgardt:1987du}
\begin{eqnarray}
{\rm Br}(\mu \rightarrow e \gamma)  &<& 2.4 \times 10^{-12},
\label{Brmu2e_exp} \\
{\rm Br}(\mu \rightarrow 3e )  &<& 1.0 \times 10^{-12},
\label{Brmu23e_exp}
\end{eqnarray}
respectively, while the EDM of the electron is very suppressed
~\cite{Hudson:2011zz}
\begin{eqnarray}
|d_e| &<&1.05 \times 10^{-27}~ e~{\rm cm}.
\label{de_exp}
\end{eqnarray}
Another important reason is that the chirality flip in the radiative
processes ({\it i.e.}, $\mu \rightarrow e \gamma$ and electron EDM) is
not suppressed by the masses of the muon and electron respectively
since mass mixing of the heavy leptons in a loop diagram can cause the
chirality flip.  Regarding $\mu \rightarrow 3e$, we will see that this
process occurs at tree level, which enhances its importance compared
with the loop process $\mu \rightarrow e\gamma$. This is only
partially degraded by the phase space suppression associated with the
additional particle in the final state.

The standard model leptons are quite light and so if there are no
awkward cancellations in their mass matrices one expects hierarchical
dimensionless standard model Yukawa couplings $Y_e^{ij}$ in the range,
$10^{-2}-10^{-5}$. Naively one might expect that the same is true for
both the couplings $\lambda_E^i$ and $\lambda_L^{i}$ since the fields
$L_L^{\prime}$ and $e_R^{\prime}$ have the same quantum numbers as the
standard model fields. As will be shown later, the parts of the
amplitudes for $\mu \rightarrow e \gamma$ and electron EDM that are
not suppressed by the muon or electron mass are proportional to the
product of dimensionless couplings, $ \epsilon=\lambda_E \lambda_L
h_E^{\prime,\prime \prime}$. (Here the flavor indices of
$\lambda_{L,E}^i$ are omitted.) Hence, the above argument would give
rise to the expectation that $\epsilon$ is very small, ${\it i.e.}$,
in the range $10^{-4}-10^{-10}$. However this is not necessarily the
case. To make the argument quantitative we now show that it is
possible to find a global $U(1)$ symmetry that forces the standard
model Yukawa coupling constants $Y_e^{ij}$ to vanish but forces only
one of the couplings in the product $\epsilon$ to vanish.  Under this
symmetry the fields transform as: $L_L^i \rightarrow e^{i q_L \alpha}
L_L^i$, $e_R^i \rightarrow e^{i q_e \alpha} e_R^i$, $e_1 \rightarrow
e^{i\ q_1\alpha}e_1$ and $e_2 \rightarrow e^{i q_2\alpha}e_2$. We take
$q_e \ne q_L$ to forbid the standard model Yukawa interactions. Then
it is not possible to choose the charges $q_1$ and $q_2$ so all the
coupling constants $\lambda_E^{i}$, $\lambda_L^{i}$ and $h_E^{\prime,
  \prime \prime}$ are non-zero. But it is possible to choose values of
the charges so only one of these couplings is zero,
\begin{itemize}
\item[{\it i)}] $q_1=q_e$, $q_2=q_L$ forces $h_E^{\prime}=h_E^{\prime
    \prime}=0$ but allows non zero values for $\lambda_E^{i}$ and
  $\lambda_L^{i}$.
\item[{\it ii)}] $q_1=q_2=q_L$ forces $\lambda_E^{i}=0$ but allows non
  zero values for $\lambda_L^i$ and $h_E^{\prime, \prime \prime}$.
\item[{\it iii)}] $q_1=q_2=q_e$ forces $\lambda_L^{i}=0$ but allows non
    zero values for $\lambda_E^i$ and $h_E^{\prime, \prime \prime}$.
\end{itemize}
Given these results, it is perfectly reasonable to consider
the product of couplings $\epsilon$ in the range $10^{-2}-10^{-5}$ and
perhaps even larger if one takes a different point of view than we
have adopted.

We now transition to the mass eigenstate basis for the charged lepton
fields which we denote by a tilde. The $5\times5$ charged lepton mass
matrix $\cal M$ is defined by
\begin{eqnarray}
{\cal L}_{\rm mass} &=& - 
\begin{pmatrix}
\bar{e}_1 & \bar{e}_2 & \bar{e}^i 
\end{pmatrix}_L
\begin{pmatrix}
M_L   &  m'     & \mu^i_E \\
m''^* &  M_E    & 0_{1\times 3} \\
0_{3\times1}     & \mu^i_L & 0_{3\times 3} 
\end{pmatrix}
\begin{pmatrix}
e_1 \\
e_2 \\
e^i
\end{pmatrix}_R
+h.c.
\nonumber \\
&\equiv& - \bar{E}_L {\cal M} E_R,
\end{eqnarray}
where $E=(e_1,e_2, e^i)^T$. Here we take $M_L$ and $M_E$ real and
define $m'=h'_E v$, $m''=h''_E v$, $\mu^i_E=\lambda^i_E v$,
$\mu^i_L=\lambda^i_L v$, where $v$ is the vacuum expectation value of
Higgs field. For our purposes we can set the standard model Yukawa
couplings $Y_e^{ij}$ to zero.  ${\cal M}$ is diagonalized  using
two unitary matrices as follows
\begin{eqnarray}
{\cal M}_{\rm diag} = V_L {\cal M} V_R^{\dagger},
\end{eqnarray}
where the mass eigenstates,
$\tilde{E}=(\tilde{e}_1,\tilde{e}_2,\tilde{e}^i)$, are given by
\begin{eqnarray}
\tilde{E}_R &=& V_R E_R ,
\label{ER}
\\
\tilde{E}_L &=& V_L E_L .
\label{EL} 
\end{eqnarray}
Thus for $i=1,2,3$ the $\tilde{e}^i$ are the familiar $e,\mu,\tau$.
We derive $V_R$ and $V_L$ by treating off-diagonal elements in ${\cal
  M}$ as a perturbation.  More explicitly we assume that $v/M_{E,L}\ll
1$ and that $|M_L-M_E|\gg v$ and perform perturbation theory in the
small quantity $v/M_{E,L}$. At the leading non-trivial order of
perturbation theory,
\begin{eqnarray}
V_R &=&
\begin{pmatrix}
1      & -p_R  & q_R^i \\
p_R^*  &  1    & 0_{1 \times 3} \\
-q_R^{* i}&  0_{3 \times 1}    & 1_{3 \times 3}  
\end{pmatrix},
\label{VR}
\\
V_L &=&
\begin{pmatrix}
1      & - p_L  & 0_{1 \times 3} \\
p_L^*  &  1     & q_L^{* i}\\
0_{3 \times 1}      &  -q_L^i  & 1_{3 \times 3} 
\end{pmatrix},
\label{VL}
\end{eqnarray}
where
\begin{eqnarray}
&&
p_R = \frac{M_Lm'+M_Em''}{M_E^2-M_L^2},\ \ \ q^i_R = \frac{\mu^i_E}{M_L},
\\ &&
p_L = \frac{M_Lm''+M_Em'}{M_E^2-M_L^2},\ \ \ q^i_L = \frac{\mu^i_L}{M_E}.
\end{eqnarray}

In the neutrino sector, on the other hand, the heavy neutrino $\nu_1$
and linear combinations of ordinary neutrinos are mass eigenstates.

We will perform calculations of loop-level processes in $R_{\xi}$
gauge with $\xi=1$ where not only the contribution from the Higgs
boson $h$ but also the fictitous scalar $\phi^0$ and $\phi^+$ ({\it
  i.e.}, the neutral and charged Goldstone boson) must be included.
Using Eqs.~(\ref{ER})-(\ref{VL}) we find that their Yukawa couplings
in the mass eigenstate basis are (to the order we require)
\begin{eqnarray}
&&
{\cal L}_{\rm Yukawa} =
\nonumber \\ && 
 \frac{h}{\sqrt{2}} 
\Bigl[
-\lambda^i_E \bar{\tilde{e}}_{1L} \tilde{e}^i_R
+(h''^*_E q_R^i-\lambda^i_Ep^*_L) \bar{\tilde{e}}_{2L} \tilde{e}^i_R  
\nonumber \\
&& +(h''_Eq^{i*}_L+\lambda^{i*}_Lp_R) \bar{\tilde{e}}_{1R} \tilde{e}^i_L
-\lambda^{i*}_L \bar{\tilde{e}}_{2R} \tilde{e}^i_L
\Bigr]
\nonumber \\ &&
+  i\frac{\phi^0}{\sqrt{2}}
\Bigl[
-\lambda^i_E \bar{\tilde{e}}_{1L} \tilde{e}^i_R
-(h''^*_E q_R^i+\lambda^i_Ep^*_L) \bar{\tilde{e}}_{2L} \tilde{e}^i_R\nonumber \\
&&+(h''_Eq^{i*}_L-\lambda^{i*}_Lp_R) \bar{\tilde{e}}_{1R} \tilde{e}^i_L
+\lambda^{i*}_L \bar{\tilde{e}}_{2R} \tilde{e}^i_L
\Bigr]
\nonumber \\ &&
+\phi^+\Bigl[
-\lambda^i_E\bar{\nu}_{1L}\tilde{e}^i_R 
+ h''_E q_L^{i*}\bar{\nu}_{1R}\tilde{e}^i_L
\Bigr]
+h.c. .
\end{eqnarray}
The flavor changing couplings of the $Z$ boson to the light charged
leptons is important for the calculation of $\mu \rightarrow 3e$. The
relevant terms are
\begin{eqnarray}
&&
{\cal L}_Z = 
-\frac{g_Z}{2} Z_\mu 
\left[
q^{i*}_Rq^j_R \bar{\tilde{e}}^i_R \gamma^\mu\tilde{e}_R^j
-q^{i}_Lq^{j*}_L \bar{\tilde{e}}^i_L \gamma^\mu\tilde{e}_L^j
\right]
\nonumber \\ &&
+g_Z Z_\mu \left[
\sin^2 \theta_W \bar{\tilde{e}}_R^i\gamma^\mu \tilde{e}_R^i+
\Bigl(-\frac{1}{2}+\sin^2 \theta_W \Bigr)
\bar{\tilde{e}}_L^i\gamma^\mu \tilde{e}_L^i
\right],
\nonumber \\
\end{eqnarray}
where $g_Z=\sqrt{g_1^2+g_2^2}$ ($g_1$ and $g_2$ are gauge coupling
constants of $U(1)_Y$ and $SU(2)_L$, respectively) and $ \theta_W$ is
the Weinberg angle.  Integrating out the heavy leptons induces flavor
mixing in the $Z$ couplings to the light standard model leptons which
causes $\mu \rightarrow 3e$ at tree level.

\section{Results}
\begin{figure}[t]
  \begin{center}
    \includegraphics[scale=0.55]{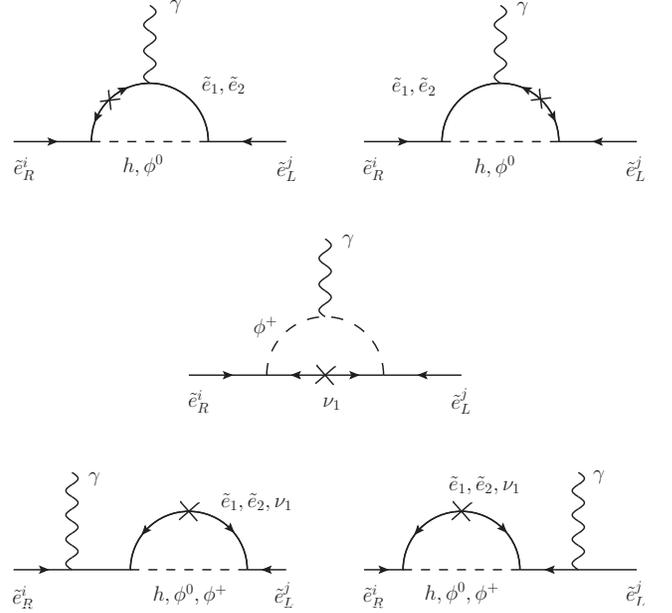}
  \end{center}
  \caption{Diagram for the $\tilde{e}^i_R \rightarrow \tilde{e}^j_L
    \gamma$. Here arrows show the chirality of fermions and chirality
    flip is indicated by a cross. }
  \label{fig:ampMuEGam}
\end{figure}

The effective Lagrangian density that describes the impact of
integrating out the heavy leptons on the light charged lepton dipole
moments and on radiative transitions between light charged lepton
flavors is,
\begin{eqnarray}
{\cal L}_{\rm eff} = 
C_{ij} \bar{\tilde{e}}^i_L \sigma^{\mu\nu} \tilde{e}^j_R F_{\mu\nu}
+C_{ji}^* \bar{\tilde{e}}^i_R \sigma^{\mu\nu} \tilde{e}^j_L F_{\mu\nu},
\end{eqnarray} 
where $\sigma^{\mu\nu}=i[\gamma^\mu,\gamma^\nu]/2$.  The coefficients
$C_{ij}$ are derived by computing the amplitude for $\tilde{e}^i_R
\rightarrow \tilde{e}^j_L \gamma$ process
(Fig.~\ref{fig:ampMuEGam}). We find that,
\begin{eqnarray}
C_{ij} = \frac{e}{64\pi^2} 
\left[
\sum_{B,I} c_{ij}^{BI} \frac{K_e(M_I,m_B)}{M_I}
-c_{ij}^{\phi^+1}\frac{K_\nu(M_1,M_W)}{M_1}
\right],
\nonumber \\
\end{eqnarray}
where $e$ is electric charge, $B=h,\phi^0$ and $I=1,2$ ({\it
  i.e.}, the index labeling heavy lepton type) and recall that
$M_1\simeq M_L$ and $M_2\simeq M_E$.

The coefficients $c_{ij}^{BI}$ are,
\begin{eqnarray}
c_{ij}^{h1}&=& \frac{1}{2}(h''^*_Eq^i_L+\lambda^i_Lp_R^*) \lambda^j_E,
\label{eq:ch1} \\
c_{ij}^{\phi^01}&=& \frac{1}{2}(h''^*_Eq^i_L-\lambda^i_Lp_R^*) \lambda^j_E,
\label{eq:cphi1}
\end{eqnarray}
for $h$-$\tilde{e}_1$ and $\phi^0$-$\tilde{e}_1$ loop diagrams, and
\begin{eqnarray}
c_{ij}^{h2}&=& \frac{1}{2}\lambda^i_L(h''^*_Eq^j_R-\lambda^j_Ep_L^*) ,
\label{eq:ch2} \\
c_{ij}^{\phi^02}&=& \frac{1}{2}\lambda^i_L(h''^*_Eq^j_R+\lambda^j_Ep_L^*) ,
\label{eq:cphi2}
\end{eqnarray}
for $h$-$\tilde{e}_2$ and $\phi^0$-$\tilde{e}_2$ loop diagrams, and
\begin{eqnarray}
c_{ij}^{\phi^+1} = h''^*_Eq^i_L \lambda^j_E,
\end{eqnarray}
for $\phi^+$-$\nu_1$ loop diagrams.  $K_e(M_I,m_B)$ and
$K_\nu(M_1,M_W)$ are loop function given by
\begin{eqnarray}
K_e(M_I,m_B)&=&\frac{1}{(1-r)^3}
\Bigl[
(1-r)(1-3r)-2r^2\log r
\Bigr],
\nonumber \\ \\
K_\nu(M_1,M_W)&=& \frac{1}{(1-r)^3}
\Bigl[
1-r^2+2r\log r
\Bigr],
\end{eqnarray}
where $r=m_B^2/M_I^2$ and $r=M_W^2/M_1^2$ respectively. $K_{e,\nu}
\simeq 1$ when $r\ll 1$.  Since we computed in $R_{\xi}$ gauge with
$\xi=1$ the appropriate mass for the fictitious scalars are
$m_{\phi^0}=M_Z$ and $m_{\phi^+}=M_W$. In this gauge the contribution
from the $Z$ and $W$ bosons are negligible. For numerical results, we
will take $m_h=125~{\rm GeV}$~\cite{Aad:2012tfa,Chatrchyan:2012ufa}.
In the region where $M_{L,E}$ are multi-TeV, the Wilson coefficients
$C_{ij}$ are given by the very simple expression,
\begin{eqnarray}
C_{ij} \simeq 
\frac{e\lambda^i_L\lambda^j_Em''^*}{64\pi^2M_EM_L}.
\label{Cij}
\end{eqnarray}
Here we have neglected contributions to the Wilson coefficients
$C_{ij}$ that are suppressed by factors of the light lepton masses.

\begin{figure}[t]
  \begin{center}
   \includegraphics[scale=0.55]{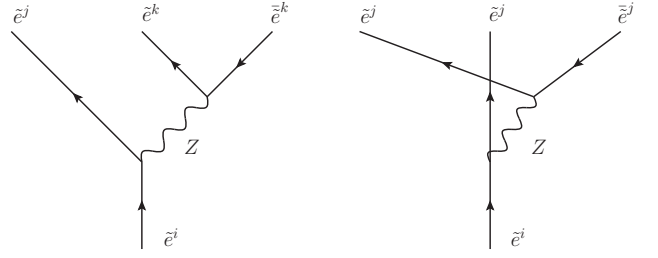}
  \end{center}
  \caption{Diagram of $\tilde{e}^i \rightarrow
    \tilde{e}^j\tilde{e}^k\bar{\tilde{e}}^k$ process. When $k\neq j$,
    only left diagram is relevant, meanwhile both diagrams contribute
    when $k=j$.}
  \label{fig:ampMuEEE}
\end{figure}

The process $\tilde{e}^i \rightarrow
\tilde{e}^j\tilde{e}^k\bar{\tilde{e}}^k$ is induced at tree level
(Fig.~\ref{fig:ampMuEEE}). The decay rate of this process is 
given by (assuming $m_i\gg m_j,m_k$),
\begin{eqnarray}
\label{etoeee1}
&&
\Gamma(\tilde{e}^i \rightarrow
\tilde{e}^j\tilde{e}^j\bar{\tilde{e}}^j) =
\frac{m_i^5}{1536\pi^3}
\nonumber \\ &&
\times \left[
\Bigl|\frac{\lambda_E^i\lambda_E^j}{M_L^2}\Bigr|^2 (\kappa^2_L+2\kappa^2_R)
+\Bigl|\frac{\lambda_L^i\lambda_L^j}{M_E^2}\Bigr|^2 (2\kappa^2_L+\kappa^2_R)
\right],
\nonumber \\
\end{eqnarray}
for $k=j$, where $\kappa_L=-1/2+\sin^2 \theta_W$,
$\kappa_R=\sin^2 \theta_W$, and 
\begin{eqnarray}
\label{etoeee2}
\Gamma(\tilde{e}^i \rightarrow
\tilde{e}^j\tilde{e}^k\bar{\tilde{e}}^k) =
\frac{(\kappa^2_L+\kappa^2_R)m_i^5}{1536\pi^3}
\left[
\Bigl|\frac{\lambda_E^i\lambda_E^j}{M_L^2}\Bigr|^2 
+\Bigl|\frac{\lambda_L^i\lambda_L^j}{M_E^2}\Bigr|^2 
\right],
\nonumber \\
\end{eqnarray}
for $j\neq k$.

The radiative processes which we are interested in are obtained from
the Wilson coefficients $C_{ij}$. The decay rate for
$\tilde{e}^i\rightarrow \tilde{e}^j \gamma$ is given by
\begin{eqnarray}
\Gamma(\tilde{e}^i\rightarrow \tilde{e}^j \gamma)
= \frac{1}{4\pi}(|C_{ij}|^2+|C_{ji}|^2) m_i^3,
\label{mu2e}
\end{eqnarray}
where $m_i$ is the mass of $\tilde{e}^i$ and we have assumed $m_i \gg
m_j$.

EDMs $d_i$ of the charged leptons, on the other hand, are coefficients
of the effective Lagrangian
\begin{eqnarray}
{\cal L}_{\rm EDM} = - \frac{i}{2} d_i
\bar{\tilde{e}}^i \sigma^{\mu\nu}\gamma_5\tilde{e}^i F_{\mu\nu},
\end{eqnarray}
and so $d_i = -2{\rm Im}(C_{ii})$.  For later discussion we also give
anomalous magnetic dipole moments (MDM) $a_i$ of the charged leptons,
which are similarly given by
\begin{eqnarray}
{\cal L}_{\rm MDM} =  \frac{e}{4m_{i}} a_i
\bar{\tilde{e}}^i \sigma^{\mu\nu}\tilde{e}^i F_{\mu\nu},
\end{eqnarray}
implying that  $ a_i =  (4m_i/e){\rm Re}(C_{ii})$. 

To summarize,  we find that
\begin{eqnarray}
\label{mainresultsa}
\Gamma(\tilde{e}^i\rightarrow \tilde{e}^j \gamma) &=&
\frac{\alpha v^2 m_i^3|h^{\prime \prime}_E|^2
(|\lambda_L^i\lambda_E^j|^2+|\lambda_L^j\lambda_E^i|^2)}
{4096\pi^4M_E^2M_L^2},
\nonumber \\
 \\
d_i &=& 
-\frac{e v}{32\pi^2}
\frac{{\rm Im}(h_E^{\prime \prime *}\lambda^i_L\lambda^i_E)}{M_E M_L},
\\  \label{mainresultse}
a_i &=& 
\frac{m_i v}{16\pi^2}
\frac{{\rm Re}(h_E^{\prime \prime *}\lambda^i_L\lambda^i_E)}{M_E M_L}.
\label{mainresultsf}
\end{eqnarray}
Eqs.~(\ref{etoeee1}), (\ref{etoeee2}) and
(\ref{mainresultsa})-(\ref{mainresultsf}) are the principal analytic
results of this paper. They are valid in the limit that $v/M_{L,E}\ll
1$ and $m_{i}/v\ll 1$

Plugging in the values of the standard model parameters and focusing
on the quantities ${\rm Br}(\mu \rightarrow e \gamma$), ${\rm Br}(\mu
\rightarrow 3e$), $|d_e|$ and $|a_{\mu}|$ the above implies that,
\begin{eqnarray}
\label{target1}
{\rm Br}(\mu \rightarrow e \gamma)&\simeq&
2.3 \times 10^{-12}\left|{h_E^{\prime \prime} \over 10^{-4}}\right|^2
\left( {10^2 ~{\rm TeV^2} \over M_EM_L} \right)^2 \nonumber \\
&&\times (| \lambda_L^{\mu}\lambda_E^e|^2+| \lambda_L^e\lambda_E^{\mu}|^2),
\end{eqnarray}

\begin{eqnarray}
\label{target2}
{\rm Br}(\mu \rightarrow 3e)&\simeq&
1.7 \times 10^{-8} 
\nonumber \\ &&  \! \! \! \! \! \! \! \! \! \! \! \! \! \! \! \! 
\! \! \! \! \! \! \! \! \! \! \! \! \! \! \! \! \! \! \! \! 
\times
\left[ 
\Bigl| \lambda^e_E\lambda^\mu_E{10^2 ~{\rm TeV^2} \over M_L^2} \Bigr|^2
+ 1.1\times
\Bigl| \lambda^e_L\lambda^\mu_L{10^2 ~{\rm TeV^2} \over M_E^2} \Bigr|^2
 \right],
\nonumber \\
\end{eqnarray}
\begin{equation}
\label{target3}
|d_e| \simeq 1.1\times 10^{-26} e~{\rm cm} 
\left|{\rm Im}\left(
{h_E^{\prime \prime *}\lambda^e_L\lambda^e_E \over 10^{-4}}
\right)\right|\left( {10^2 ~{\rm TeV^2} \over M_EM_L} \right),
\end{equation}
and
\begin{equation}
\label{target4}
|a_{\mu}|\simeq  1.2\times 10^{-13}
\left|{\rm Re}\left(
{h_E^{\prime \prime *}\lambda^{\mu}_L\lambda^{\mu}_E \over 10^{-4}}
\right) \right|\left( {10^2 ~{\rm TeV^2} \over M_EM_L} \right).
\end{equation}
Eqs. (\ref{target1}), (\ref{target2}), (\ref{target3}) and
(\ref{target4}) are compared with the measured values given in
Eqs.~(\ref{Brmu2e_exp}), (\ref{Brmu23e_exp}) and (\ref{de_exp}) and the
deviation of the observed muon $g-2$ ~\cite{Bennett:2006fi} from its
standard model prediction~\cite{Davier:2010nc},
\begin{eqnarray}
\Delta a_\mu^{\rm exp} = a_\mu^{\rm exp}-a_\mu^{\rm SM} &\simeq&
(2.8\pm0.8)\times 10^{-9}.
\end{eqnarray}
Note that while it is expected that a 10 TeV vector-like lepton can
give an observable contribution to $\mu \rightarrow e \gamma$, $\mu
\rightarrow 3e$ and the electron EDM, such a heavy vector-like lepton
doesn't give a large enough contribution to the anomalous magnetic
moment to explain the discrepancy between standard model value and its
observed value.

Before going on to the numerical results, we comment on the
perturbative corrections to light lepton mass matrix that come from
integrating out the heavy leptons.  There is a tree-level correction
to light lepton mass matrix elements, which is induced by mixing with
heavy leptons. It is given by
\begin{eqnarray}
|\Delta m_{ij}^{\rm tree}| = |m'' q^i_Rq^j_L|.
\label{mijTree}
\end{eqnarray}
In addition to this correction, there is also a logarithmically
divergent one loop correction to the light lepton mass matrix that
arises from one loop self-energy diagrams analogous to the Feynman
diagrams that contribute to the radiative transitions and dipole
moments but without the photon. Using a momentum cutoff $\Lambda$, it
is
\begin{eqnarray}
&&\Delta m_{ij}^{\rm loop} 
= -\frac{1}{16\pi^2}\sum_{B,I}c_{ij}^{BI} 
M_I {\rm log}\left({\Lambda^2 \over M_I^2} \right) \nonumber \\
&&\simeq -\frac{\lambda_L^i\lambda_E^j m^{\prime \prime *}}{16\pi^2} 
\left[{2M_L \over M_E}{\rm log}\left({\Lambda^2 \over  M_L^2}\right)
+{M_E \over M_L}
   {\rm log}\left({\Lambda^2 \over M_E ^2}\right)\right],
\nonumber \\
\label{mijLoop}
\end{eqnarray}
assuming $\Lambda \gg M_I\gg m_h,m_Z$. This quantum correction is more
important than the tree-level one when $M_{L,E}$ is greater than about
a few hundred GeV. In the region of heavy vector-like leptons masses
we are focusing on the tree level contribution from mixing with the
heavy leptons can be neglected and so in order to avoid awkward
cancellations between $\Delta m_{ij}=\Delta m_{ij}^{\rm tree}+ \Delta
m_{ij}^{\rm loop} \simeq \Delta m_{ij}^{\rm loop}$ and the standard
model tree level light lepton mass matrix elements from
Eq.~(\ref{smyuk}) (which are proportional to $Y_e^{ij}$), we adopt the
naturalness criteria~\cite{Cheng:1987rs}:
\begin{eqnarray}
|\Delta m_{ij}\Delta m_{ji}|\lesssim m_i m_j.
\label{finetune}
\end{eqnarray}
A more precise analysis of naturalness constraints would involve
solving renormalization group equations for the evolution of the
coupling constants in the theory

It is important to remember that the region of parameter space
excluded by Eq.~(\ref{finetune}) is not ruled out and can be
consistent with experiment. We view the region of parameter space that
does not satisfy Eq.~(\ref{finetune}) as disfavored by the naturalness
criteria that the one loop correction to the light lepton mass matrix
calculated with a large momentum space cutoff not overwhelm the
contribution which is coming from Eq.~(\ref{smyuk}).

We have computed the parts of the amplitudes for $\mu \rightarrow e
\gamma$ and the electric dipole moment of the electron that are not
suppressed by the light lepton masses.  The radiative correction to
the light lepton mass matrix in Eq.~(\ref{mijLoop}) is proportional to
the same combination of couplings constants that occur in the
amplitudes for these radiative processes.  Hence the naturalness
condition in Eq.~(\ref{finetune}) restricts the size of this part of
their amplitudes. There are situations where the radiative correction
to the charged lepton mass matrix is negligible and the rate for $\mu
\rightarrow 3e$ not suppressed, for example, if $h_E^{\prime \prime}$
is extremely small but $\lambda_{E,L}$ order unity. In this case the
contributions to the amplitudes for the radiative processes
proportional to the light lepton masses, that we have not calculated,
dominate. However, in that situation the rate for $\mu \rightarrow 3e$
is sensitive to higher mass scales for the vector-like leptons than
the rate for $\mu \rightarrow e \gamma$ since ${\rm Br}(\mu
\rightarrow e \gamma) \sim (\alpha/4\pi){\rm Br}(\mu \rightarrow 3
e)$.

\begin{widetext}

\begin{figure}[t]
  \begin{center}
    \includegraphics[scale=0.6]{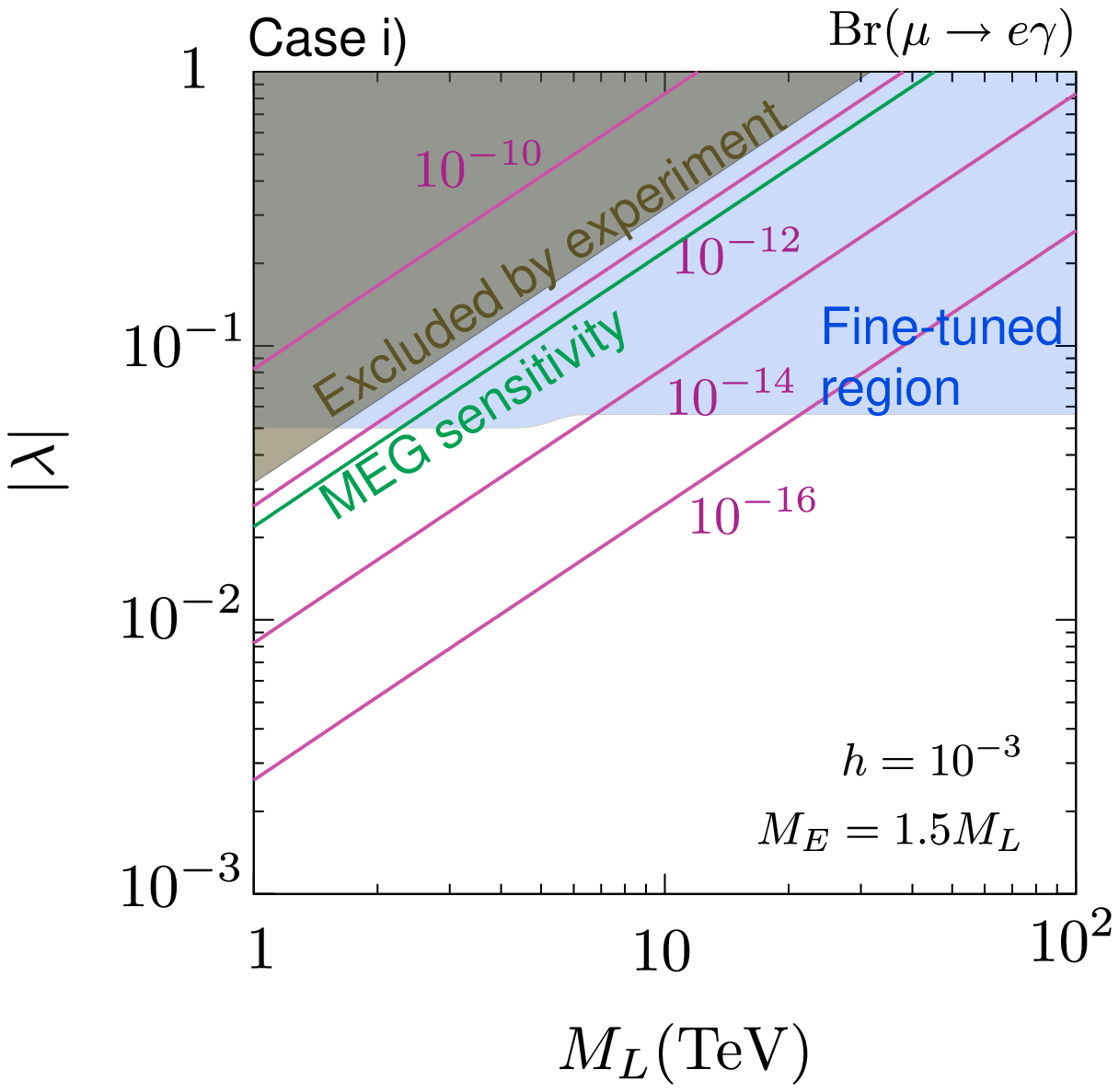} \! \! \! \! \! \!
    \includegraphics[scale=0.6]{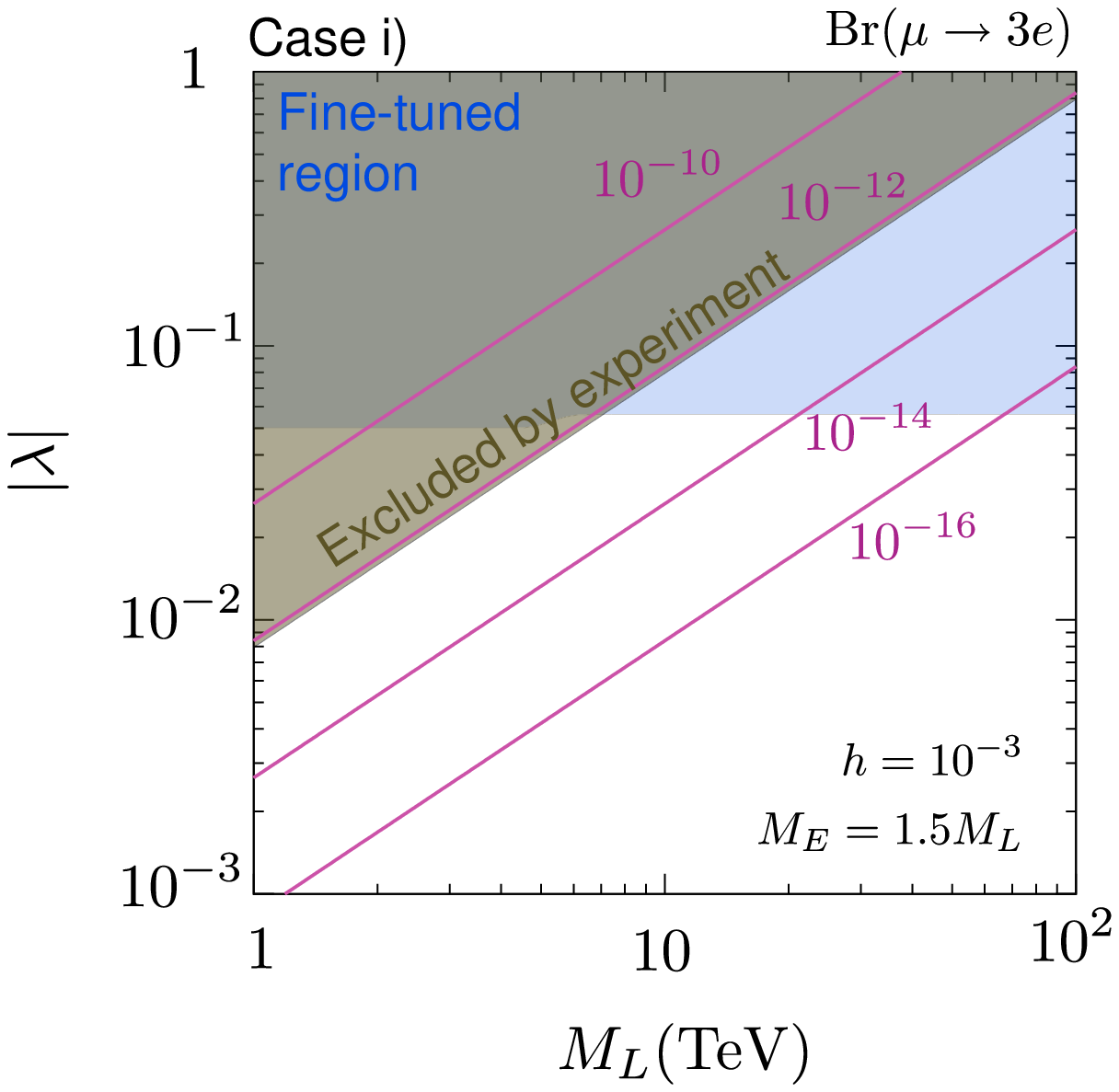}
    \includegraphics[scale=0.6]{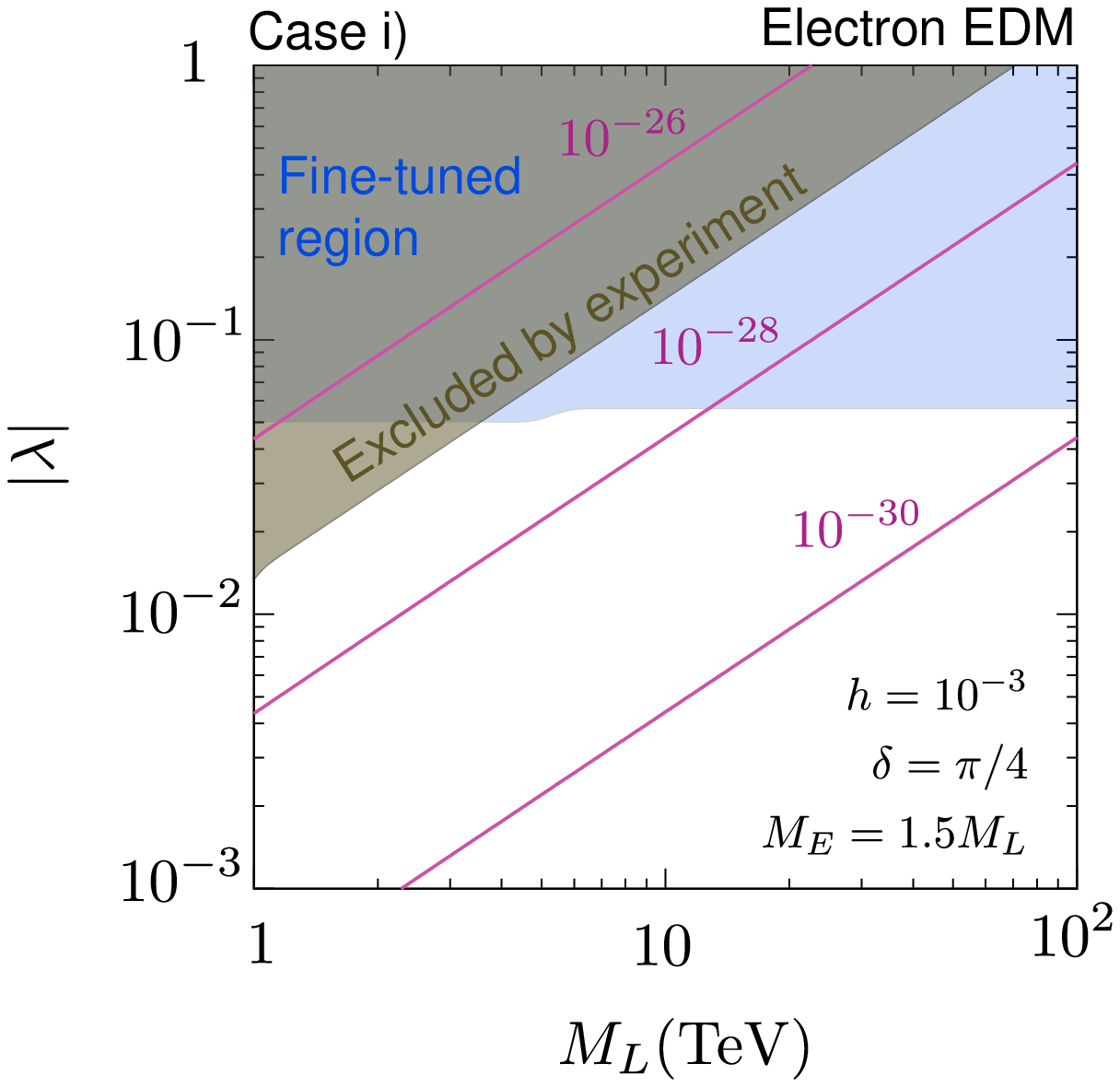} \! \! \! \! \! \!
    \includegraphics[scale=0.6]{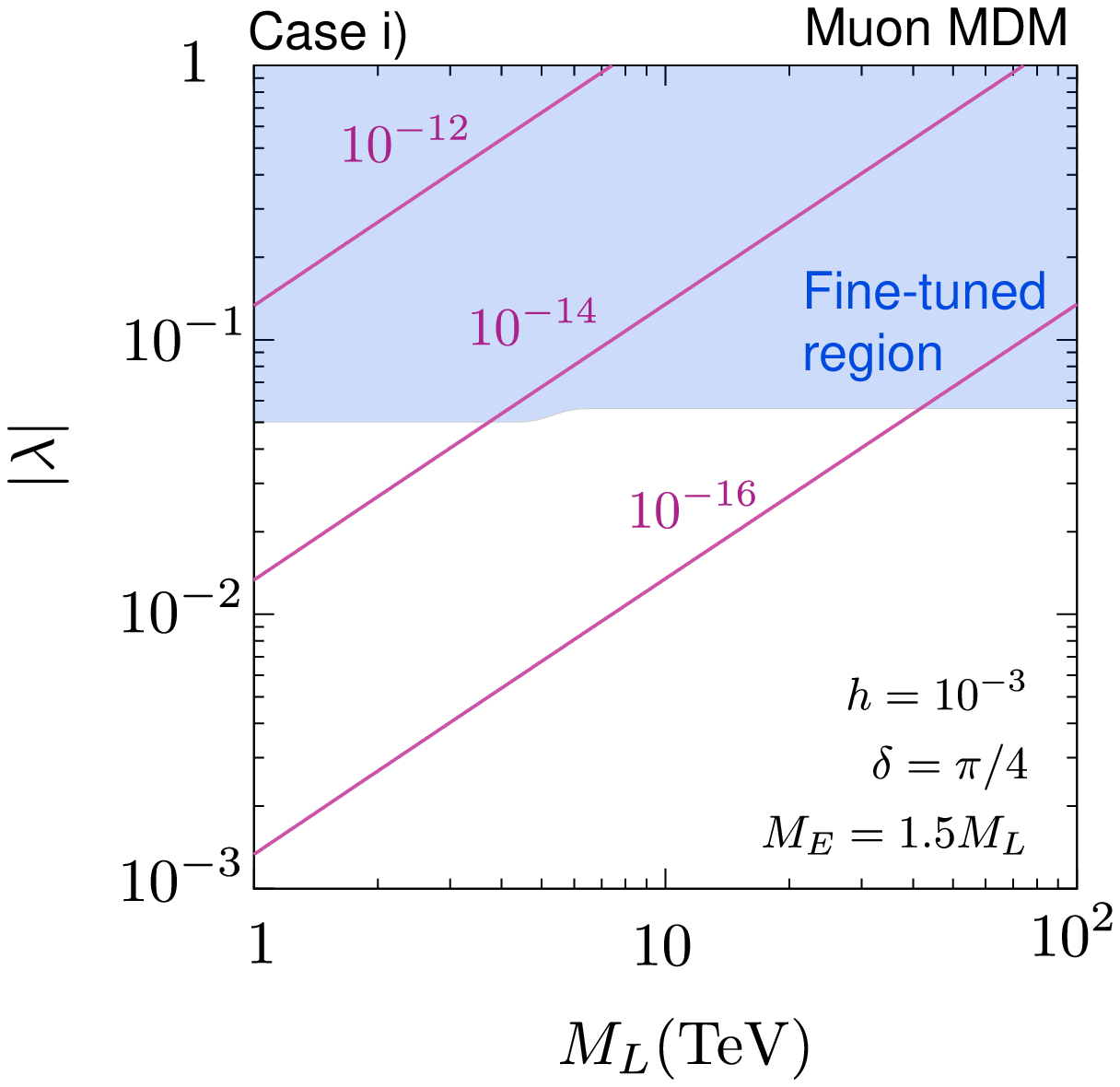}
  \end{center}
  \caption{Contours of fixed ${\rm Br}(\mu \rightarrow e \gamma)$ (top
    left), ${\rm Br}(\mu \rightarrow 3e)$ (top right), electron EDM
    (bottom left) and muon MDM (bottom right) in the $M_L$-$|\lambda|$
    plane for the parameters discussed in the text.  Lines are ${\rm
      Br}(\mu \rightarrow e \gamma)={\rm Br}(\mu \rightarrow
    3e)=10^{-10},10^{-12},10^{-14},10^{-16}$,
    $|d_e|=10^{-26},10^{-28},10^{-30}~e~{\rm cm}$, and
    $|a_\mu|=10^{-12},10^{-14},10^{-16}$ from top to
    bottom. The dark-shaded region is excluded by experiments, while
    light-shaded (blue) indicates fine-tuned parameter space.  In the
    top left, the prospective MEG sensitivity is also plotted. }
  \label{fig:case1}
\end{figure}
\end{widetext}

\section{Numerical Results}

First we consider case {\it i)} discussed in Sec.~\ref{sec:model}. We
take $h'_E=h''_E=h=10^{-3}$, $\lambda_E^i=\lambda_L^i=\lambda$ and
$M_E=1.5M_L$.  Fig.~\ref{fig:case1} shows contours of fixed ${\rm
  Br}(\mu \rightarrow e \gamma)$, ${\rm Br}(\mu \rightarrow 3e)$,
electron EDM and muon MDM in the $M_L$-$|\lambda|$ plane. The
dark-shaded region is excluded by experiments.  We also shade the
fine-tuned parameter space where Eq.~(\ref{finetune}) is not
satisfied, taking $\Lambda = 10^{15}~{\rm GeV}$. Note that, $\mu
\rightarrow 3e$ gives a more stringent constraint on the parameter
space. This is because ${\rm Br}(\mu \rightarrow 3e)$ is not
suppressed by $h$.  The current experimental bound constrains
$|h\lambda^2|\lesssim 10^{-5}$ when $M_{L}\sim 10~{\rm TeV}$. On the
other hand, the region of parameter space where $|h\lambda^2|\gtrsim
10^{-5}$ is fine-tuned.  For the electron EDM, we take $\delta=\pi/4$
where ${\rm arg}(\lambda^2)=\delta$.  The region of the plane excluded
by experiment is almost the same as in the $\mu \rightarrow 3e$
case. Note that the rate for $\mu \rightarrow 3e$ does not depend on
$h$ and if we take $h$ much smaller then the fine-tuned region
disappears.

There are also experimental constraints on flavor changing $\tau$
lepton decays. Since the weak decay width of a charged lepton of type
$i$ is proportional to $m_i^5$, the branching fractions for $\tau
\rightarrow \mu \gamma$ and $\tau \rightarrow e \gamma$ satisfy ${\rm
  Br}(\tau \rightarrow \mu \gamma)\simeq {\rm Br}(\tau \rightarrow e
\gamma)\simeq (m_{\mu}/m_{\tau})^2{\rm Br}(\mu \rightarrow e \gamma)$
in  case {\it i)}. Current experimental bounds for these process
are by ${\rm Br}(\tau \rightarrow \mu \gamma) < 4.4\times 10^{-8}$ and
${\rm Br}(\tau \rightarrow e \gamma) < 3.3\times 10^{-8}$
~\cite{Aubert:2009ag}. Thus these bounds are less stringent than for
muon decays.

For reference, we have given the contours of fixed muon MDM. Clearly
the impact of vector-like leptons on the muon MDM is negligible in the
parameter region we have focused on. In order to explain the muon $g-2$
anomaly by vector-like leptons, they should be lighter than 1 TeV, for
example, $M_{L,E}\sim O(100~{\rm GeV})$ with $|h\lambda^2 \cos
\delta|\sim 10^{-3}$.  Note that having vector-like leptons with
masses around the weak scale doesn't necessarily lead to a conflict
with the limits on the rates for $\mu \rightarrow e \gamma$, $\mu
\rightarrow 3e$ and the electron electric dipole moment.  For example,
it could be the case that $|\lambda_{E,L}^e|$ is much smaller than
$|\lambda_{E,L}^\mu|$.

\begin{widetext}

\begin{figure}[t]
  \begin{center}
   \includegraphics[scale=0.6]{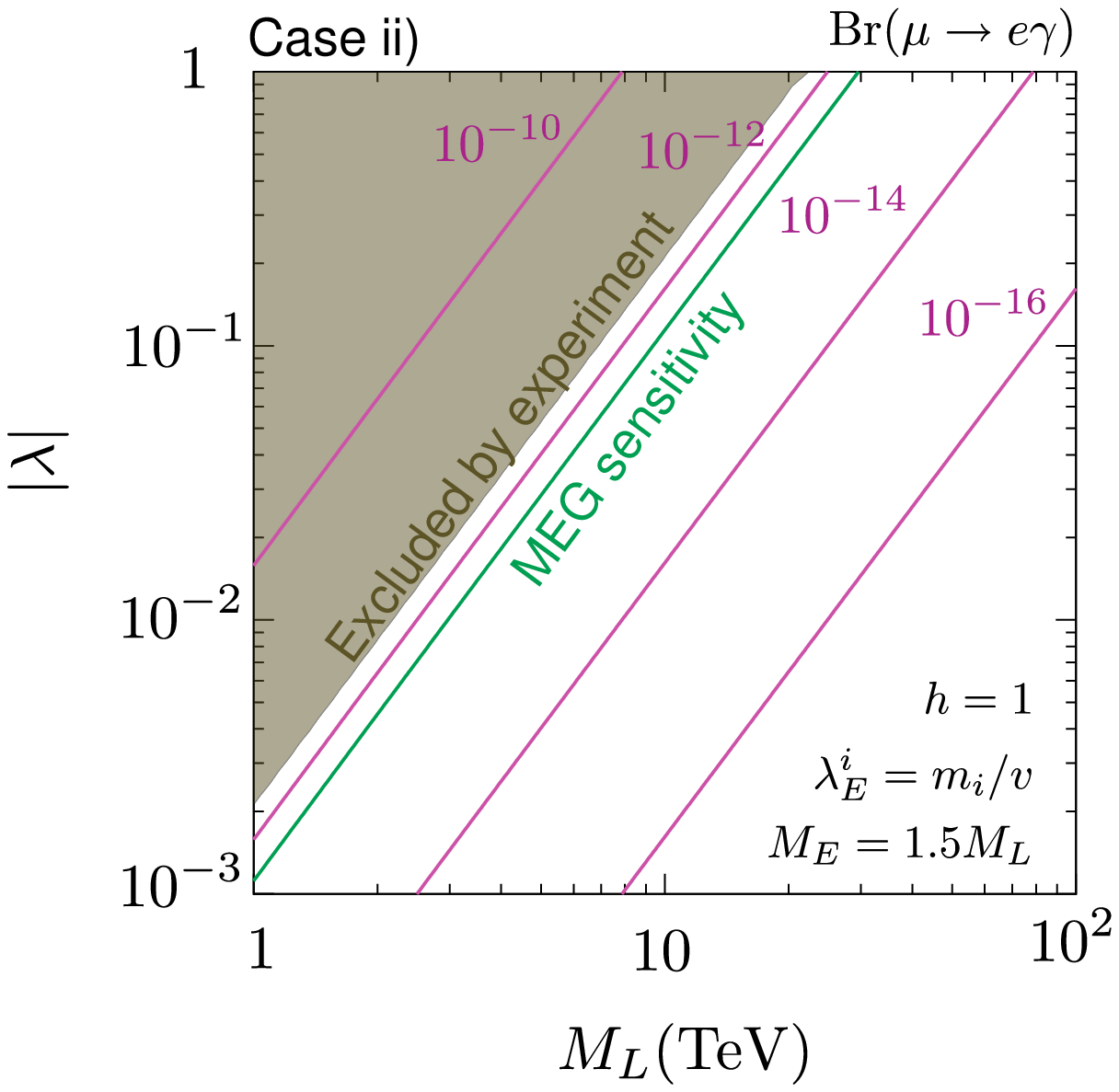} \! \! \! \! \! \!
   \includegraphics[scale=0.6]{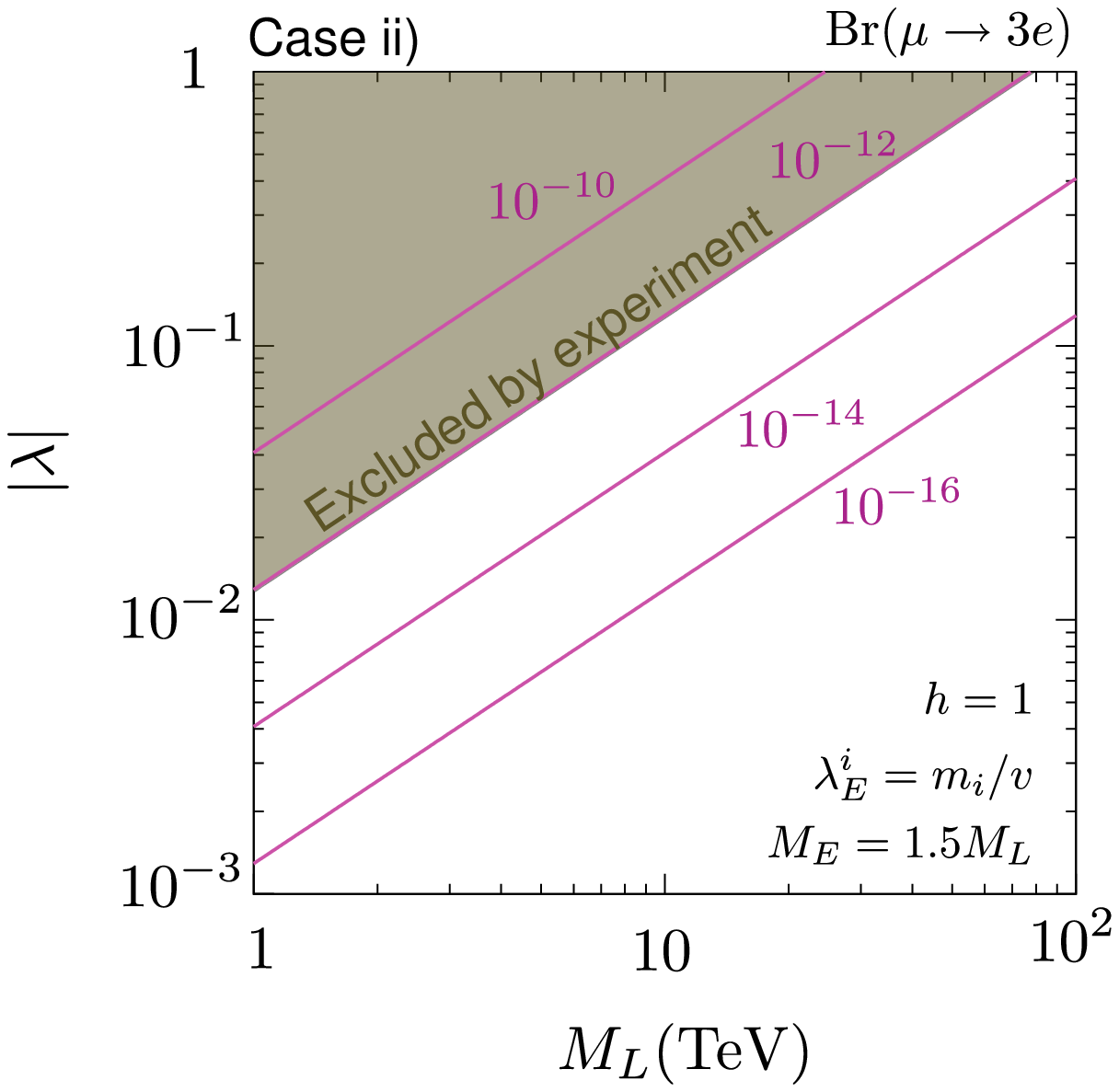} 
   \includegraphics[scale=0.6]{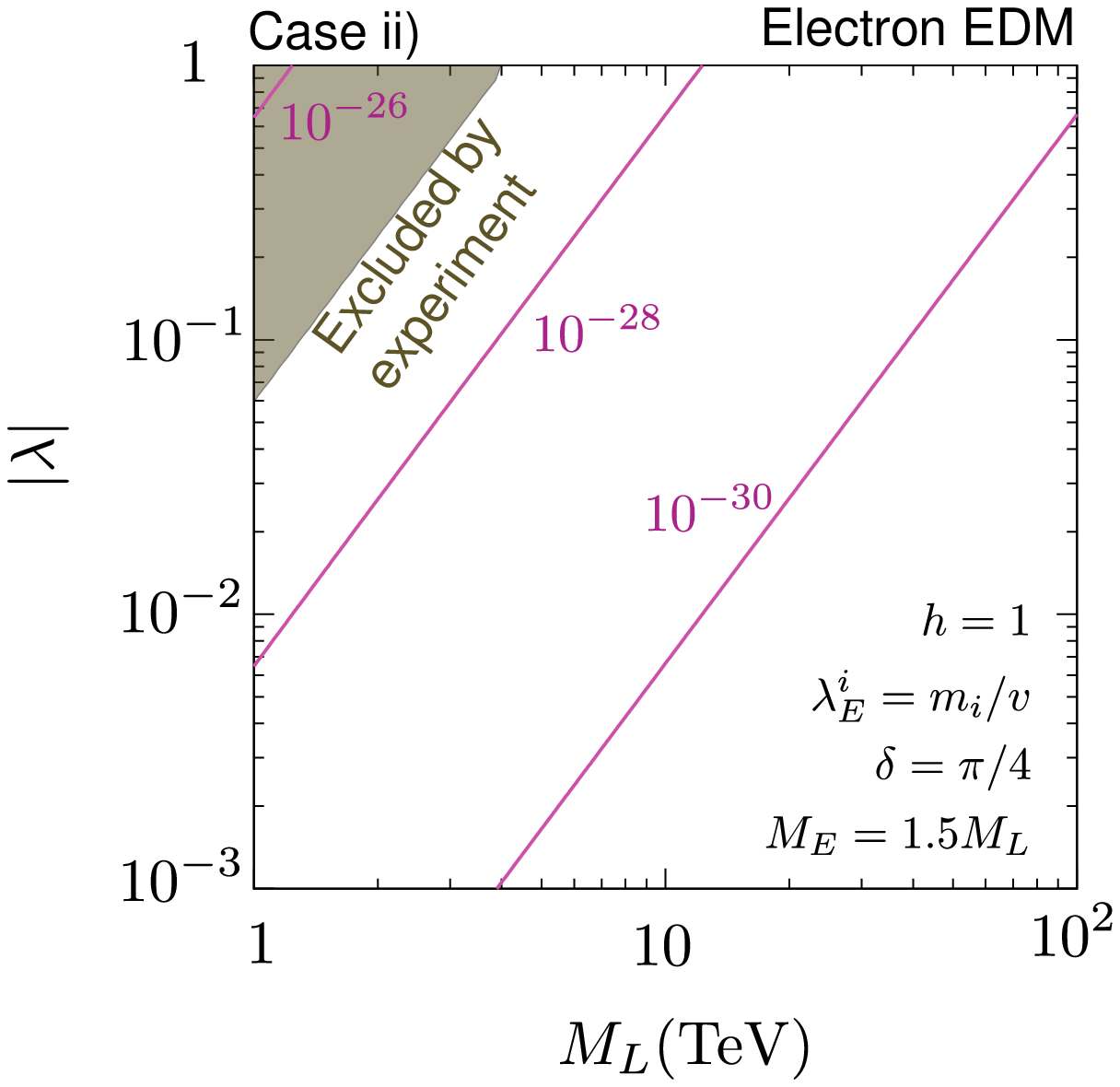}\! \! \! \! \! \!
    \includegraphics[scale=0.6]{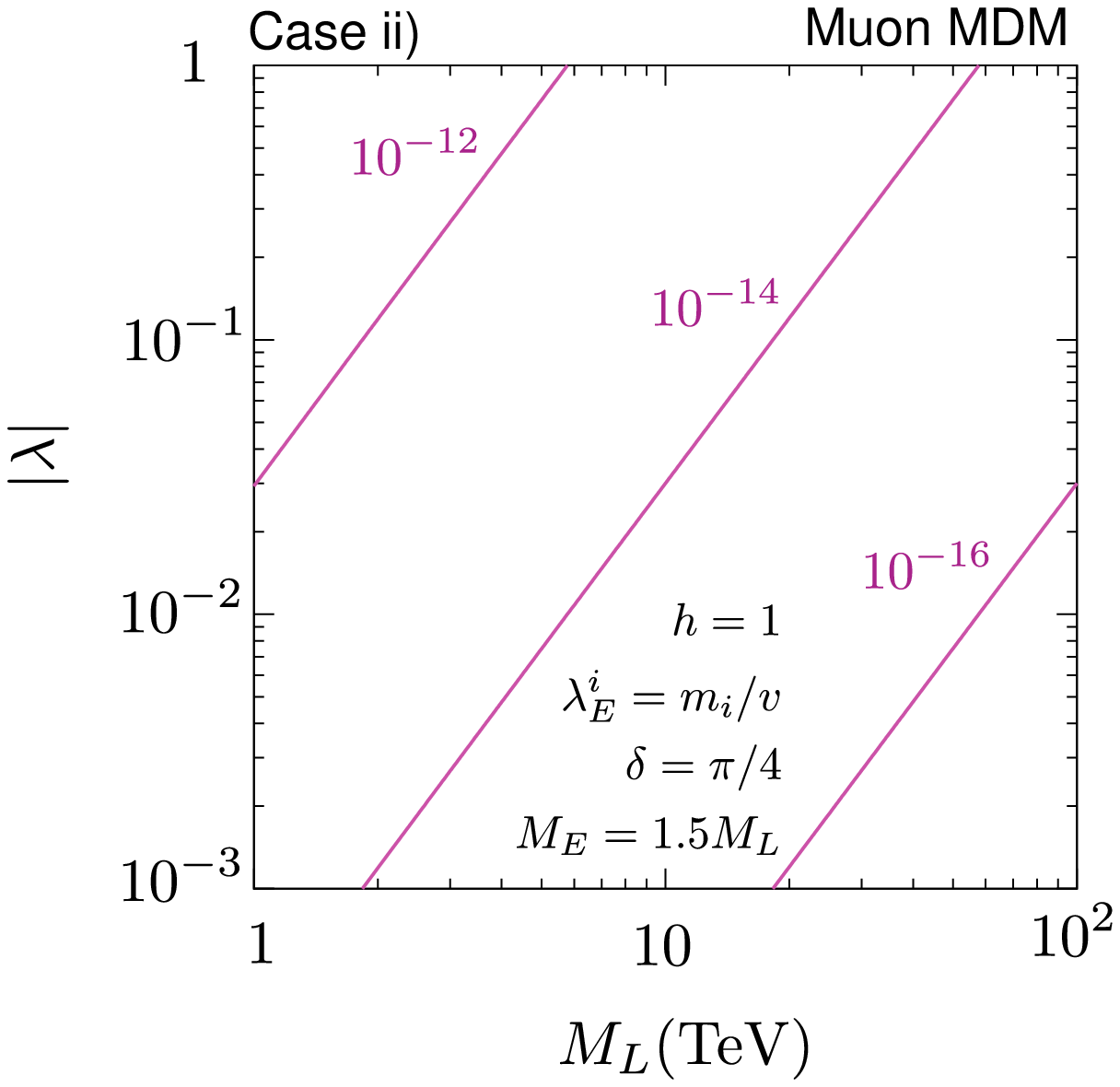}
  \end{center}
  \caption{ Same as Fig.~\ref{fig:case1} but with different choice
    of parameters.}
  \label{fig:case2}
\end{figure}

\end{widetext}

Next we consider the case {\it ii)}. (Parametrically case {\it iii)}
is identical to case {\it ii)}.) We take $h'_E=h''_E=h=1$,
$\lambda^e_L=\lambda^\mu_L=\lambda$, $\lambda_E^i=m_i/v$ and
$M_E=1.5M_L$.  Fig.~\ref{fig:case2} shows the results. (For the
electron EDM, we take $\delta={\rm arg}(\lambda)=\pi/4$.) In this case
there is no fine-tuned region (taking $\Lambda=10^{15}~{\rm
  GeV}$). Since the electron EDM is suppressed by $|\lambda_E^e|$, the
experimental bound becomes weaker compared to the previous case.  The
current bounds from the limits on the $\mu \rightarrow e \gamma$ and
$\mu \rightarrow 3e$ branching ratios, on the other hand, give a more
severe constraint on the parameter space. For $\mu \rightarrow e
\gamma$, this is because the branching ratio is only suppressed by
$|\lambda_E^\mu|$.  The numerical result indicates that the
$M_{E,L}\sim 10-100~{\rm TeV}$ region may be probed in the future MEG
experiment, which has prospective sensitivity of ${\rm Br}(\mu
\rightarrow e \gamma)\simeq 5.0\times 10^{-13}$. (For $\tau$ radiative
decay, we note that ${\rm Br}(\tau \rightarrow \mu \gamma)={\rm
  Br}(\mu \rightarrow e \gamma)$ in this case.)  The process $\mu
\rightarrow 3e$, on the other hand, is not suppressed by any of the
$|\lambda^i_E|$ (see Eq.~(\ref{target2})), which gives rise to large
branching fraction. It is especially interesting that the $M_{L,E}\sim
100~{\rm TeV}$ region with $|\lambda|\sim O(1)$ is already excluded by
current experiments, which means vector-like leptons with masses up to
100 TeV can have an observable impact on experiments. 

\section{Concluding Remarks}
We have considered the phenomenological impact of vector-like leptons
that are much heavier than the weak scale. We discussed expectations
for their couplings based on the observed masses of the charged
leptons. With couplings that are reasonable from this perspective we
found that vector-like leptons with masses up to $100~ {\rm TeV}$ can
have an impact on the electric dipole moment of the electron and the
rates for $\mu \rightarrow 3e$ and $\mu \rightarrow e \gamma$ that is
detectable in the next generation of experiments. When the vector-like
leptons are this heavy their impact on the anomalous magnetic moment
of the muon and Higgs properties is very small.

For the chirality flipping radiative processes $\mu \rightarrow e
\gamma$ and the electric dipole moment of the electron we have
focused on the parts of their amplitudes that are not suppressed by
the light lepton masses. One of the goals of this work has been to
explore whether the absence of light lepton mass suppression results
in sensitivity to much higher masses for the vector-like leptons. We
found that one-loop radiative corrections to the light lepton mass
matrix are proportional to the same combination of couplings and
demanding that the one loop contribution not overwhelm the
contribution from Yukawa couplings restricts the reach in vector-like
lepton mass. Finally we noted that, for similar branching ratios, the
three body flavor changing decay $\mu \rightarrow 3e$ has sensitivity
to greater vector-like lepton masses than the radiative two body decay
$\mu \rightarrow e \gamma$.

\subsection*{Acknowledgment}
This paper is funded by the Gordon and Betty Moore Foundation through
Grant $\# 776$ to the Caltech Moore Center for Theoretical Cosmology
and Physics. The work of the authors was supported also in part by the
U.S. Department of Energy under contract No. DE-FG02-92ER40701. The
research of MBW was supported in part by Perimeter Institute for
Theoretical Physics. Research at Perimeter Institute is supported by
the Government of Canada and by the Province of Ontario through the
Ministry of Economic Development \& Innovation.


\end{document}